\documentclass[aps, prd, twocolumn,print, showpacs, nofootinbib]{revtex4}
\usepackage{graphicx}

\begin{document}

\title{High Redshift Gamma-Ray Bursts: Observational Signatures of
Superconducting Cosmic Strings?}
\author{K. S. Cheng$^1$}
\author{Yun-Wei Yu$^{1,2}$}
\author{T. Harko$^1$}

\affiliation{$^1$Department of Physics, The University of Hong Kong,
Pokfulam
Road, Hong Kong, China\\
$^2$Institute of Astrophysics, Huazhong Normal University, Wuhan,
China}

\begin{abstract}
The high-redshift gamma-ray bursts (GRBs), GRBs 080913 and 090423,
challenge the conventional GRB progenitor models by their short
durations, typical for short GRBs, and their high energy releases,
typical for long GRBs. Meanwhile, the GRB rate inferred from
high-redshift GRBs also remarkably exceeds the prediction of the
collapsar model, with an ordinary star formation history. We show
that all these contradictions could be eliminated naturally, if we
ascribe some high-redshift GRBs to electromagnetic bursts of
superconducting cosmic strings. High-redshift GRBs could become a
reasonable way to test the superconducting cosmic string model,
because the event rate of cosmic string bursts increases rapidly
with increasing redshifts, whereas the collapsar rate decreases.
\end{abstract}


\date{\today}

\maketitle


{\it Introduction}---Cosmological gamma-ray bursts (GRBs; see
\cite{Pir04} for reviews and references therein) are usually
classified phenomenologically into two classes, i.e., the
long-duration, soft-spectrum class, and the short duration,
hard-spectrum class \cite{Kou93}. An observer-frame duration $\sim2$
s is traditionally taken as the separation line. For long GRBs,
their host galaxies are typically irregular (in a few cases spiral)
galaxies with intense star formation and, especially, a handful of
long GRBs are firmly associated with Type Ib/c supernovae. So it is
nearly confirmed that most (if not all) of long GRBs are produced
during the core collapse of massive stars (called collapsars). In
contrast, short GRBs are usually found at nearby early-type
galaxies, with little star formation. All deep supernova searches
for them have led to nondetections. These facts are in good
agreement with the conjecture that short GRBs could originate from
mergers of compact binaries. Although this simple classification
looks so solid, it is challenged by a few unusual GRBs, most
notably, the two highest-redshift GRBs (GRBs 080913 and 090423, with
$z=6.7$ \cite{hzgrb1} and $8.2$ \cite{hzgrb2}, respectively). On one
hand, both GRBs 080913 and 090423 have an intrinsic duration shorter
than 2 s, and a hard spectrum, typical for short GRBs. On the other
hand, they are also quite unlikely to be short GRBs, in view of
their high energy releases, and their consistency with the Amati
relation \cite{ama02}, which is only fulfilled by long GRBs. These
contradictions make these high-redshift GRBs very difficult to be
understood in both progenitor models of collapsars and mergers of
compact binaries.

It is usually suggested to use long GRBs as cosmological tools to
probe the star formation history of the Universe. The rapidly
increasing number of GRBs with known redshifts has motivated several
such attempts, under the assumption that the GRB rate traces the
star formation rate (SFR), either with a constant ratio
\cite{cha07}, or more probably with an additional evolution
\cite{yuk08,kis08}. Using a sample including GRBs 080913 and 090423,
and assuming all of the high-redshift GRBs are collapsar GRBs, the
SFR-determination was extended to a redshift interval never explored
before (only $\sim630$ Myr after the Big Bang) \cite{Kis09}. While
the SFR at $4<z<5$ is in basic agreement with earlier measurements,
the SFR inferred from the high-redshift ($z>6$) GRBs seems to be too
high in comparison with the one obtained from some high-redshift
galaxy surveys \cite{Ho06,Bou08} (although the missing of the faint
end of the galaxy luminosity function could also suppress the
corresponding estimations \cite{Bou08}). The overestimation of the
high-redshift SFR from the GRB counts naturally makes us to consider
the possibility that the high-redshift GRB sample could be seriously
polluted by a fair number of non-collapsar GRBs, which correspond to
a new, intrinsically different GRB origin. In the very early
Universe, superconducting cosmic strings could be the most plausible
candidates.

Cosmic strings are linear topological defects that could be formed
at a symmetry breaking phase transition in the early Universe, as
predicted in most grand unified models. Strings can respond to
external electromagnetic fields as thin superconducting wires
\cite{Wi85}, and thus be able to develop electric currents as they
move through cosmic magnetic fields. Therefore, due to the
oscillating loops of the superconducting strings, highly beamed,
short electromagnetic bursts would be emitted from  small string
segments, centering at some peculiar points (i.e., cusps), where
the velocity nearly reaches the speed of light \cite{Wi86, Os86}.
An observational evidence for such cosmic string bursts has
probably been provided by an observed millisecond radio burst
\cite{lor07}, as argued by Vachaspati \cite{vac08}. The
possibility that strings can serve as GRB engines was first
suggested in \cite{Ba87,Pa88} and further carefully investigated
in \cite{Br93,Be01}. The powerful electromagnetic radiation from a
cusp eventually produces a jet of accelerated particles, whose
internal dissipations and terminational shock are possibly
responsible for the GRB prompt and afterglow emissions,
respectively. Before the {\it Swift} era, it was almost impossible
to distinguish cosmic-string GRBs (CSGRBs) from conventional
collapsar GRBs at relatively low redshifts, because the event rate
of the CSGRBs decreases dramatically with the expansion of the
Universe. Presently, such a situation may be essentially changed
by the discovery of GRBs with very high redshifts.


{\it GRBs from superconducting cosmic strings}---At any time $t$, a
horizon-size volume contains a few long strings stretching across
the volume, and a large number of small scale wiggles. The typical
distance between the long strings and their characteristic curvature
radius are both on the order of $\sim ct$, while the wavelengthes of
the wiggles are down to $l\sim \alpha ct$, where $c$ is the speed of
light. In this Letter we treat, for simplicity, the value of
$\alpha$ as a constant, as it is usually done, although in principle
it could  decrease with time due to string radiation \cite{vac08}.
The exact value of $\alpha$ is not known, but numerical simulations
still gave an upper bound, $\alpha\lesssim10^{-3}$, while a lower
bound $\alpha\gtrsim50G\mu/c^2\sim5\times10^{-9}$ is determined by
the gravitational radiation backreaction \cite{Be01} with $G$ being
the Newton's constant and $\mu\sim10^{18}\rm g~cm^{-1}$ being the
mass per unit length of string. The string can be treated as a
series of closed loops that oscillate with a period $T_l\sim l/2c$.
The number density of the loops can be estimated as
$n_l\sim1/\alpha^{}c^{3}t^{3}$. Oscillating loops tend to form
cusps, where the string segment rapidly reaches a speed very close
to the speed of light. Near a cusp, the string gets contracted by a
large factor, and its rest energy is converted into kinetic energy.

As shown by Witten \cite{Wi85}, strings predicted in a wide class of
elementary particle models behave as superconducting wires. If an
electric field $E$ is present on the axis of a superconducting
string, the string generates an electric current $I$ at the rate
$dI/dt\sim (ce^2/\hbar) E$, where $e$ is the electron charge. Then,
for a string segment moving with velocity $\sim c$ in an external
magnetic field $B$, we can get the current increase rate as
$dI/dt\sim (ce^2/\hbar)B$. Hence a superconducting loop oscillating
in a magnetic field can act as an alternating current generator,
developing a current of amplitude $I_0\sim (e^2/\hbar)Bl$
\cite{Wi86}. Due to the large current, a powerful electromagnetic
radiation is generated, and the emitted power can be estimated with
the use of the magnetic dipole radiation formula as $P_0\sim
{m^2\omega^4/ c^3}\sim {I_0^2/ c}$ \cite{Wi86},
where $m\sim I_0l^2/c$ is the magnetic moment of the loop and
$\omega\sim T_l^{-1}$ is the typical frequency of the oscillation.
Taking the direction of the string velocity at the cusp as
$\theta=0$ and considering the radiation from all segments nearby
the cusp, the radiation from the angle $\theta$ would be mainly
contributed by the segment with Lorentz factor
$\gamma\sim\theta^{-1}$. As a result, the angular distribution of
the total power of the loop is given by ${dP/d\Omega }\sim
kP_0/\theta^3$ \cite{Wi86}, where $k\sim10$.

For a viewing angle $\theta$ with respect to the string velocity at
the cusp, the radiation received by observers is from the string
segment with Lorentz factor $\gamma\sim\theta^{-1}$. Then the
observational luminosity would be determined by \cite{Ba87,foot}
\begin{equation}
L_{ }\sim\gamma^{3}{ dP\over d\Omega }\sim kP_0\gamma^6
\sim10^{52}{\rm
erg~s^{-1}}~\alpha_{-8}^2B_{0,-7}^{2}\gamma_{3}^{6}f_{z,1}^{},\label{lum}
\end{equation}
where $f_z\equiv(1+z)$ and hereafter the convention $Q_x=Q/10^{x}$
is adopted for the cgs units. The duration of this radiation in the
local inertial frame can be estimated to be $\tau_{\rm loc}\sim
T_l/\gamma$. Considering the Doppler effect and the cosmological
time dilation further, the duration as seen by observers can be
calculated by
\begin{eqnarray}
\tau&\sim&\tau_{\rm loc}(1-\beta)f_z\sim
{T_{l}\over\gamma^{3}}f_z\sim1~{\rm
s}~\alpha_{-8}\gamma_{3}^{-3}f_{z,1}^{-1/2},\label{Tauc1}
\end{eqnarray}
where $\beta=(1-\gamma^{-2})^{1/2}$. See reference \cite{Ba87} for a
detailed discussion on this timing estimation. In the above
calculations, the time is approximated by
$t\approx(1/H_0)f_z^{-3/2}$, with $H_0= 73~ \rm km~s^{-1}Mpc^{-1}$,
and the magnetic field, assumed to be frozen in the cosmic plasma,
is simply calculated as $B(z)=B_0f_z^2$ with $B_0$ being the
magnetic field strength at the present time. From Eq.~(\ref{lum}),
the Lorentz factor can be inferred from the observational
luminosities as
\begin{eqnarray}
\gamma\sim10^3~\alpha_{-8}^{-1/3}B_{0,-7}^{-1/3}L_{ 52}^{1/6}f_{z,1}^{-1/6}.\label{gamma}
\end{eqnarray}
Substituting this expression into Eq.~(\ref{Tauc1}) and using
$\tau=T_{90}$ (the observational duration $T_{90}$ is the period in
which 90 percent of the burst's energy is emitted), the main model
parameters can be constrained by
\begin{eqnarray}
\alpha_{-8}^{2}B_{0,-7}^{}\sim9~T_{90,1}L_{ 52}^{1/2},\label{abrel1}
\end{eqnarray}
which is independent of the unknown $\gamma$ or $\theta$.


{\it Implications from the observed high-redshift GRBs}---There are
three types of sites in the Universe where magnetic fields can
induce sufficiently large electric currents in the strings -
galaxies and clusters of galaxies, voids, and walls (filaments and
sheets). In this Letter CSGRBs are assumed to mainly come from the
walls, which occupy a fraction of $f_B\sim 0.1$ of the space. Then
the event rate of the CSGRBs with luminosities $>L_{ }$ can be
estimated by \cite{Be01}
\begin{eqnarray}
\dot{R}_{{\rm CS}}(>L_{ })&\sim&{\theta^2\over4}f_{B}{n_l\over
T_{l} }\nonumber\\
&\sim&740~{\rm Gpc^{-3}yr^{-1}}~\alpha_{-8}^{-4/3}B_{0,-7}^{2/3}L_{
52}^{-1/3}f_{z,1}^{19/3},\label{CSRate}
\end{eqnarray}
where $\theta\sim\gamma^{-1}$ can be determined by
Eq.~(\ref{gamma}). The dependence of $\dot{R}_{{\rm CS}}$ on
$\alpha$ shows that, if a possible decrease of $\alpha$ with time is
taken into account, the increase of $\dot{R}_{\rm CS}$ with
increasing redshift would become slower. Anyway, for the {\it Swift}
satellite, with an angular sky coverage of $\Delta \Omega/4\pi\sim
0.1$, and for a five-year observation period ($\mathcal T\sim 5$
yr), the expected number of CSGRBs between redshifts $z$ and
$z+\Delta z$ can be calculated as \cite{Be01}
\begin{eqnarray}
\mathcal N^{\rm exp}_{\rm CS}(>L_{ })&=&\mathcal P\mathcal T{\Delta
\Omega\over 4\pi}\int_{z}^{z+\Delta z}\dot{R}_{\rm
CS}f_{z'}^{-1}dV_p(z'),\label{csnumber}
\end{eqnarray}
where the factor $f_{z'}^{-1}$ is due to the cosmological time
dilation of the observed rate, and $0<\mathcal P<1$ indicates the
ability both to detect the initial burst of gamma rays, and to
obtain a redshift from the optical afterglow. The proper volume
element is given by \cite{Be01} $dV_p(z')=\left[4\pi d_c^2
(c/H_0)(f_{z'}^3\Omega_m+\Omega_\Lambda)^{-1/2}dz'\right]/f_{z'}^{3}$
with the comoving distance given by
$d_c=(c/H_0)\int_0^{z'}(f_{z''}^3\Omega_m+\Omega_\Lambda)^{-1/2}dz''$.
The cosmological parameters are $\Omega_m=0.27$ and
$\Omega_{\Lambda}=0.73$, respectively. Considering that the two
highest-redshift GRBs (GRBs 080913 and 090423) are CSGRBs due to
their unusual properties, the integral in Eq.~(\ref{csnumber}) over
$6.5<z<8.5$ leads to the constraint $\mathcal N^{\rm exp}_{\rm
CS}(>L_{ })\sim 15~\alpha_{-8}^{-4/3}B_{0,-7}^{2/3}L_{
52}^{-1/3}\mathcal P\geq 2$. Combing this result with
Eq.~(\ref{abrel1}), we can further obtain
$\alpha\leq4\times10^{-8}~T_{90,1}^{1/4}$, and
$B_0\geq7\times10^{-8}~{\rm G}~L_{ 52}^{1/2}T_{90,1}^{1/2}$,
which is close to the equipartition magnetic field
strength of $\sim10^{-7}$ G \cite{Ry98}.

Finally, let us return to the question asked at the beginning of
this Letter - how the SFR estimation by the high-redshift GRB sample
is influenced by the CSGRBs. Following \cite{yuk08, kis08, Kis09},
we define an effective SFR  $\dot{\rho}_{*}^{\rm eff}$, due to the
CSGRBs, as
\begin{eqnarray}
{\mathcal N^{\rm exp}_{\rm CS}(>L_{ })\over \mathcal N_{1-4}^{\rm
obs}(>L_{ })}={\int_{z}^{z+\Delta z} \xi\dot{\rho}_{*}^{\rm eff}
f_{z'}^{-1}dV_c(z') \over
\int_{1}^{4}\xi\dot{\rho}_*f_{z'}^{-1}dV_c(z')}\label{rateeqn},
\end{eqnarray}
where $\xi\propto f_{z'}^\delta$ with $\delta\sim1.5$ giving the
fraction of stars that produce GRBs, and additional evolution
effects. Different from the CSGRB rate $\dot{R}_{\rm CS}(z')$ that
is defined at the time of $t$, the SFR $\dot{\rho}_{*}(z')$ here is
measured at the present time $t_0$, so the integral of
$\dot{\rho}_{*}(z')$ should be over the comoving volume as
$dV_{c}(z')=f_{z'}^{3}dV_{p}(z')$ \cite{yuk08, kis08, Kis09}. For
the observed GRBs within $1<z<4$, we still consider that the
overwhelming majority of them originate from collapsars. So the
observational counts $\mathcal N_{1-4}^{\rm obs}(>L_{ })$ can be
estimated by integrating following star formation history
\cite{Ho06} over $1<z<4$:
\begin{equation}
\dot{\rho}_*(z)\propto\left\{
\begin{array}{ll}
f_z^{3.44},&z<0.97,\\
f_z^{-0.26},&0.97<z<4.48,\\
f_z^{-7.8},&4.48<z.
\end{array}\right.\label{sfr}
\end{equation}
with $\dot{\rho}_*(0)=0.02~{\rm M_{\odot}yr^{-1}Mpc^{-3}}$. The
actual values of $\mathcal N_{1-4}^{\rm obs}(>L_{ })$ for different
luminosities can be found from Fig.~3 in \cite{Kis09}. For
convenience, here we fit the data by $\mathcal N_{1-4}^{\rm
obs}(>L_{ })\sim15 L_{ 52}^{-\beta}$ with $\beta\sim0.78$.
Substituting Eq.~(\ref{csnumber}) into Eq.~(\ref{rateeqn}), we can
obtain the effective SFR as
\begin{eqnarray}
\dot{\rho}_*^{\rm eff}&=&{ \mathcal
T\int_{1}^{4}\dot{\rho}_*f_{z'}^{\delta-1} {dV_c}\over
f_z^{\delta}\mathcal N_{1-4}^{\rm obs}(>L_{ \rm th})}{\Delta
\Omega\over
4\pi}\mathcal P{{\dot{R}_{\rm CS}(>L_{ \rm th}) \over f_z^3}}=\nonumber\\
&& A ~{\rm M_{\odot}yr^{-1}Mpc^{-3}}~f_{z,1}^{3-\delta+\beta}
\left({f_z^{1/2}-1\over2}\right)^{2(\beta-1/3)},\label{effsfr}
\end{eqnarray}
where the prefactor $A \sim
0.025~\alpha_{-8}^{-4/3}B_{0,-7}^{2/3}\mathcal
F_{-8}^{\beta-1/3}\mathcal P$. The luminosity threshold at redshift
$z$ can be calculated as $L_{ \rm th}=4\pi d_l(z)^2\mathcal F$ for a
given flux sensitivity $\mathcal F$ and a luminosity distance
$d_l=f_zd_c\approx (3c/H_0)f_z^{1/2}(f_z^{1/2}-1)$ \cite{Be01}. With
different values of $A$, we plot the effective star formation
history inferred from the CSGRBs in Fig.~1 using dash-dotted lines.
As shown by the solid line, the SFR data (diamonds) inferred from
the high-redshift GRBs \cite{Kis09} can be well explained by
combining Eq. (\ref{effsfr}) for $A\sim0.1$ (thick dash-dotted line)
with Eq. (\ref{sfr}) (dashed line).

\begin{figure}
\resizebox{\hsize}{!}{\includegraphics{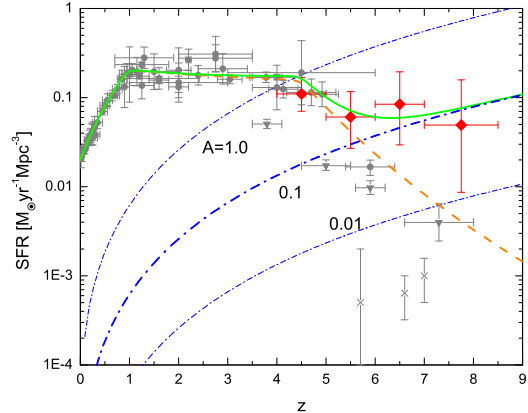}} \caption{The
cosmic star formation history. The SFR data (diamonds) inferred from
the high-redshift GRBs \cite{Kis09} obviously deviate from the star
formation history (dashed line) given by \cite{Ho06}, which is
obtained by fitting the compiled SFR data inferred from galaxies
(circles). For a comparison, we also exhibit the newer data inferred
from the high-redshift Lyman break galaxies (triangles) \cite{Bou08}
and Lyman-$\alpha$ emitters (crosses) \cite{ota08}. The dash-dotted
lines represent the effective star formation history inferred from
the CSGRBs with different values of $A$ and the solid line
represents the combination of the dash and dash-dotted line for
$A=0.1$.}
\end{figure}

{\it Conclusion and discussions}---The analysis in this Letter shows
that all of the observed luminosities, durations, and event rate of
high-redshift GRBs can be reasonably explained by ascribing some
high-redshift GRBs to superconducting cosmic string bursts.
Therefore, in the future SFR-determinations using high-redshift GRB
sample, the possible pollution from CSGRBs must be eliminated
carefully. In contrast to conventional GRBs, CSGRBs may have some
unique features, e.g., a high luminosity accompanied by a very short
duration, no association with supernova, and probably no host
galaxy. More importantly, such an ascription inevitably suggests a
new GRB original mechanism, in addition to the conventional
progenitor models as collapsars and mergers of compact binaries. In
other words, we can somewhat regard GRBs 080913 and 090423 as a new
evidence for the existence of superconducting cosmic strings in the
early Universe.

The superconducting cosmic string model may be tested in a variety
of ways. First, as mentioned above, some high-frequency (e.g., GHz)
electromagnetic waves directly radiated by a cusp at relatively low
redshift can penetrate the surrounding medium and then be detected
as a cosmic spark \cite{vac08}. Secondly, due to the quench of the
current on the strings, superconducting cosmic strings could be a
source for positrons, and thus the observed 511 keV emission from
electron-positron annihilation in the Galactic bulge can be
explained by the existence of a tangle of light superconducting
strings in the Milky Way\cite{fer05}. Thirdly, oscillating string
loops can also contribute to gravitational wave (GW) background. The
background spectrum has two main features, i.e., the ``red noise"
portion spanning the frequency range $10^{-8}$ Hz $\leq$ $f \leq
10^{10}$ Hz and the peak in the spectrum near $f\sim10^{-12}$ Hz
\cite{ViG}. Gravitational waves in the frequency band $10^{-12}$ Hz
$\leq f\leq 10^{-3}$ Hz produced by a cusp for string tensions as
small as $G\mu/c^2\sim10^{-10}$ could stand above the GW background
\cite{Da}, which might be detectable by the planned GW detectors
such as LIGO, VIRGO, and LISA.

Finally, the ascription of some high-redshift GRBs to CSGRBs enables
us to use high-redshift GRBs as a cosmological tool to constrain the
primordial cosmic magnetic fields. At present, the magnetic fields
on large scales are usually limited by the cosmic microwave
background (CMB) and by Faraday rotation measures of light from
high-redshift quasars. As a result, some upper limits (from
$10^{-9}$ G to $10^{-7}$ G \cite{Ba}) for $B_0$ have been suggested.
However, in view of the simplification of the CSGRB model and the
smallness of the high-redshift GRB sample, the numerical results in
this Letter are not yet sufficiently solid.
Anyway, the CSGRB method
at least provides a potentially effective complement to the CMB and
the Faraday rotation methods. The combination of all these methods
can give a more precise estimation on the strength of the primordial
cosmic magnetic fields.

\section*{Acknowledgements}

We thank M.C.Chu, K.M. Lee, Fa-Yin Wang and K.W. Wu for useful
discussions. KSC and TH are supported by the GRF Grants of the
Government of the Hong Kong SAR under HKU7011/09P and HKU7025/07P,
respectively. YWY is partly supported by NSFC under Grant No.
10773004.


\begin{thebibliography}{99}
\bibitem{foot}In the present Letter we consider the
bolometric luminosity only. In more details, the spectral
distribution of the released energy can be expressed by two
asymptotic functions as $\propto\exp(-\nu/\nu_{p})$ for
$\nu>\nu_{p}$ and $\propto(\nu/\nu_{p})^{-2/3}$ for $\nu<\nu_{p}$
\cite{vac08, bla01}, where $\nu$ is the observational frequency, and
$\nu_{p}\sim\gamma^3/(T_lf_z)\sim 1$ Hz (see Equation \ref{Tauc1})
is the peak frequency of the spectrum. Since this peak frequency
could be much lower than the plasma frequency of the surrounding
medium \cite{vac08}, especially in the early Universe, we simply
consider that nearly all of the total electromagnetic energy, as
expressed by the bolometric luminosity, is absorbed by the
interstellar medium to generate a GRB. Of course, by considering a
possible decay of the loops due to the radiation effect (i.e., $l$
decreases), $\nu_p$ could increase to be above the plasma frequency.
Therefore, at some relatively low redshifts, most energy may be
released as an electromagnetic wave (e.g., radio) burst \cite{vac08}
rather than to produce a GRB.
\end{thebibliography}
\end{document}